# Benchmarking Self-Driving Labs

Adedire D. Adesiji,[1,†] Jiashuo Wang,[2,†] Cheng-Shu Kuo,[3] Keith A. Brown[1,2,4,*]

[1]Department of Mechanical Engineering, Boston University, Boston, MA, USA

[2]Division of Materials Science & Engineering, Boston University, Boston, MA, USA

[3]Department of Electrical and Computer Engineering, Boston University, Boston, MA, USA

[4]Physics Department, Boston University, Boston, MA, USA

[†]These authors contributed equally

**\*Corresponding author:**

Prof. Keith A. Brown

Email: *brownka@bu.edu*

**ORCID:**

Adedire D. Adesiji: 0000-0001-8973-5748

Jiashuo Wang: 0009-0007-7223-063X

Keith A. Brown: 0000-0002-2379-2018

**Abstract:**

A key goal of modern materials science is accelerating the pace of materials discovery. Self-driving labs, or systems that select experiments using machine learning and then execute them using automation, are designed to fulfil this promise by performing experiments faster, more intelligently, more reliably, and with richer metadata than conventional means. This review summarizes progress in understanding the degree to which SDLs accelerate learning by quantifying how much they reduce the number of experiments required for a given goal. The review begins by summarizing the theory underlying two key metrics, namely acceleration factor $AF$ and enhancement factor $EF$, which quantify how much faster and better an algorithm is relative to a reference strategy. Next, we provide a comprehensive review of the literature, which reveals a wide range of $AFs$ with a median of 6, and that tends to increase with the dimensionality of the space, reflecting an interesting blessing of dimensionality. In contrast, reported $EF$ values vary by over two orders of magnitude, although they consistently peak at 10-20 experiments per dimension. To understand these results, we perform a series of simulated Bayesian optimization campaigns that reveal how $EF$ depends upon the statistical properties of the parameter space while $AF$ depends on its complexity. Collectively, these results reinforce the motivation for using SDLs by revealing their value across a wide range of material parameter spaces and provide a common language for quantifying and understanding this acceleration.

# 1. Introduction

The pace of research progress is in sharp focus due to pressing societal needs demanding the discovery of new materials.[1] The field of autonomous experimentation (AE) is addressing this challenge by developing automated systems that increase the rate and reliability of experiments while also developing algorithms that select experiments to best achieve user-defined goals.[2-4] The combination of these elements is termed a self-driving lab (SDL) (Fig. 1A), in which experiments are algorithmically selected and performed without human intervention.[5] Such systems have rapidly expanded from the first SDL for materials research less than a decade ago to now being common across materials, nanoscience, additive manufacturing, and chemistry.[6-13] The vanguard of this field has moved from demonstrations of these systems to using them for materials discoveries that have been forthcoming in areas such as lasing,[14] mechanics,[15] and battery materials.[16]

While SDLs are increasingly common, their value proposition has yet to be fully articulated, and different definitions and metrics have been proposed. Several of their virtues can be easily quantified and appreciated, such as how automation can allow additional experiments to be performed per unit time.[17, 18] A more subtle metric is how much they accelerate research, with reports ranging from 2× to 1000×.[17] One reason for this challenge is that quantifying the acceleration of research progress requires comparing the advanced strategy to some reference strategy, often necessitating additional experiments that do not directly contribute to the domain science being explored. Nevertheless, studies have established and explored different metrics that quantify the degree to which AE improves research outcomes. Two metrics that stand out are acceleration factor $AF$ and enhancement factor $EF$ that describe how much faster or better one process is relative to another (Fig. 1B).[19, 20] These metrics stand out in the context of experimental campaigns as they do not require the parameter space to be fully explored or the optimum to be known. However, comparisons are not always possible because these values are not always reported, they depend on the benchmark approach, and these metrics depend sensitively on the details of the space being explored in a manner that has not been explored for materials.

In this paper, we review the existing experimental results that benchmark the acceleration inherent to SDLs and provide insight into how to interpret these metrics. We begin by defining $EF$ and $AF$ while providing the theoretical foundation for how these should behave in a typical active learning campaign. Next, we summarize the efforts in the community to provide experimental benchmarking. Finally, we perform basic simulations that provide context for interpreting $EF$ in different parameter spaces. This review should help interpret acceleration values reported, provide guidance for the most impactful circumstances in which to apply active learning, and suggest future work in curating high-quality materials datasets for refining algorithms with direct application to materials science.

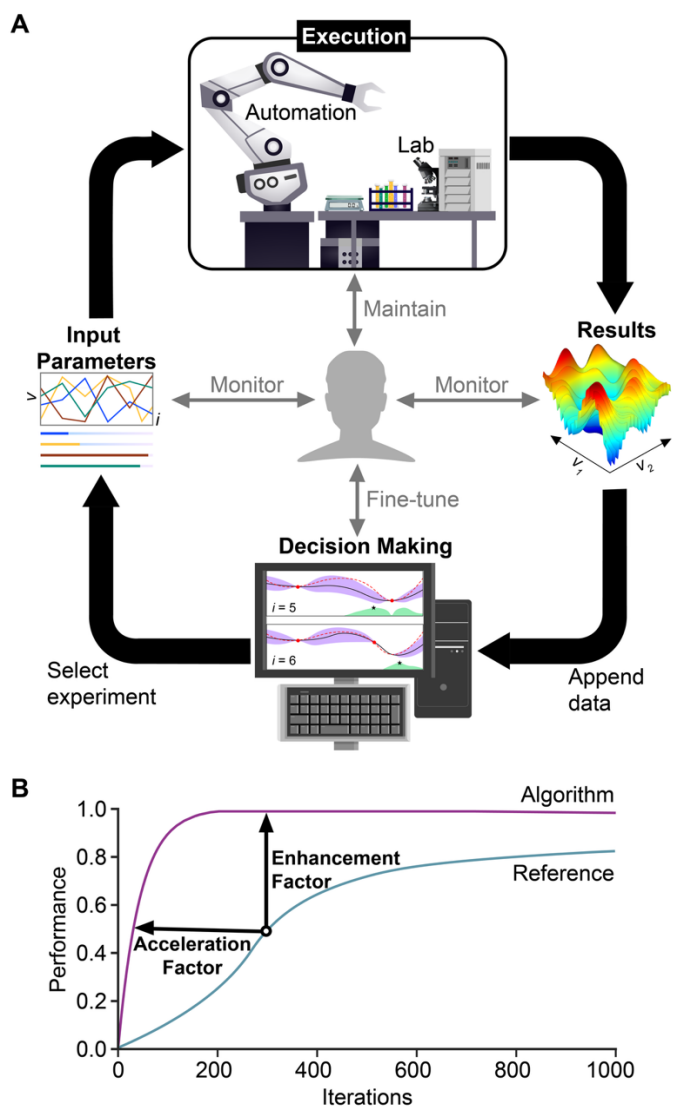

**Fig. 1** (A) Schematic of the workflow of a self-driving lab (SDL). (B) Representative performance convergence plot, also known as a horse race plot, illustrating enhancement factor *EF* and acceleration factor *AF*. *EF* quantifies relative performance after a fixed number of experiments, while *AF* quantifies the reduction in the number of experiments required to reach a target performance. Both metrics are defined relative to a reference strategy, such as sampling the space uniformly at random.

## 2. Theory

The canonical task for a materials or chemistry SDL is to run a campaign to optimize a measurable property $y$ that depends on a set of parameters $\vec{x}$. Here, $y$ can be a scalar or a vector with the latter being the purview of multi-objective optimization. Like the majority of benchmarking, we consider scalar objectives for simplicity and adopt the language of maximization, although the same logic applies to minimization tasks. The parameter space has a finite dimensionality $d$, and the variables can represent compositions, processing conditions, other conditions of the experiment, or even latent variables found using unsupervised learning. With these definitions, the goal of the campaign is to identify the conditions $\vec{x}^* = \text{argmax}(y(\vec{x}))$. After experiment number $n$ in the campaign, the progress towards this goal can be quantified by considering how close the current observed maximum $y_n^* = \max(y(\vec{x}_n))_n$ is to the true maximum $y^* = \max(y(\vec{x}))$. Interestingly, if the campaign proceeds by selecting experiments uniformly at random across $\vec{x}$, this average progress has a closed-form solution that depends upon the cumulative distribution function $F_y(y)$.[19] Specifically, the average performance after $n$ experiments corresponds to the performance at which there is a 50% chance that no better value has been observed, or

$$\frac{1}{2} = F_y(y_n)^n, \tag{1}$$

where $y_n$ is the expected best observed response from random sampling. At $n = 1$, $y_1 = \text{median}(y)$ and $y_n$ asymptotes to $y^*$ as $n \to \infty$. This simple analysis illustrates that the convergence of a simple decision-making policy depends intimately on the details of the parameter space.

While it is reasonable to derive closed-form solutions for expected convergence when the property space is known, for real materials systems, $y$ is unknowable except through experiment. The nature of continuous variables and the presence of noise in measurements mean that ground truth will never be completely known. This makes it impossible to predict how fast convergence is expected or even when the process has fully converged. Thus, benchmarking learning using an SDL involves completing two campaigns, an active learning (AL) campaign designed to test the learning algorithm along with a reference campaign guided by a standard method. From a benchmarking perspective, the most relevant data available are the best performance observed in the first $n$ experiments, defined as $y_{AL}^*(n)$ for the AL campaign and $y_{ref}^*(n)$ for the reference campaign. There are two main ways of comparing these sets of data.[19, 20] The first metric is the **acceleration factor** ($AF$) that is defined as the ratio of $n$ needed to achieve a given performance $y_{AF}$, namely,

$$AF(y_{AF}) = \frac{n_{ref}}{n_{AL}}, \tag{2}$$

where $n_{AL}$ is the smallest $n$ for which $y_{AL}^*(n) \geq y_{AF}$ while $n_{ref}$ satisfies the same condition for the reference campaign. Larger values of $AF$ indicate a more efficient AL process. The second metric is the **enhancement factor** ($EF$) that is defined as the improvement in performance after a given number of experiments, namely

$$EF(n) = \frac{y_{AL}^*(n)}{y_{ref}^*(n)}. \tag{3}$$

$EF$ presents an interesting limit when considering benchmarking using random sampling. Specifically, the very best outcome of an active learning campaign would be $y^*$ while the worst performance possible using random sampling would be median($y$), which would be the expected result at $n = 1$. This leads us to define the contrast $C$ of the property space as,

$$C = \frac{y^*}{\text{median}(y)}, \tag{4}$$

which defines the best possible $EF$ that could be found when studying that property space. Between the two metrics $EF$ and $AF$, $EF$ is often more convenient to compute as it is defined vs. $n$ and thus can be calculated at all points for reference and benchmark campaigns that have the same number of experiments.

## 3. Literature Survey

As a goal of the SDL field is accelerating progress, much work has been dedicated to benchmarking the acceleration of these systems. To comprehensively consider the literature that benchmarks active learning, we began with a broad literature search (Fig. 2). We searched the Scopus database using the keywords "Bayesian optimization" combined with "benchmark." As the field of optimization research extends far beyond its overlap with materials or SDLs, this search yielded considerable results with 4,245 publications matching these keywords (Fig. 2A). The keyword "Bayesian optimization" was chosen due to its prevalent adoption for active learning in the field of material science while the term active learning is widely used for an unrelated method in education. Most studies outside materials science utilized analytical functions or look-up tables that are designed to be challenging to optimize and thus provide insight into comparisons between learning approaches. While this broad survey is useful for evaluating active learning strategies, our focus is to evaluate benchmarking using actual experimental materials datasets. To narrow down the search to those that involved benchmarking using self-driving labs, we conducted a search with the broad term "self-driving lab", resulting in 111 studies. After examining each study, only 40% of these articles reported direct efforts to benchmark performance. These data are provided at https://github.com/kabrownlab/benchmarking.

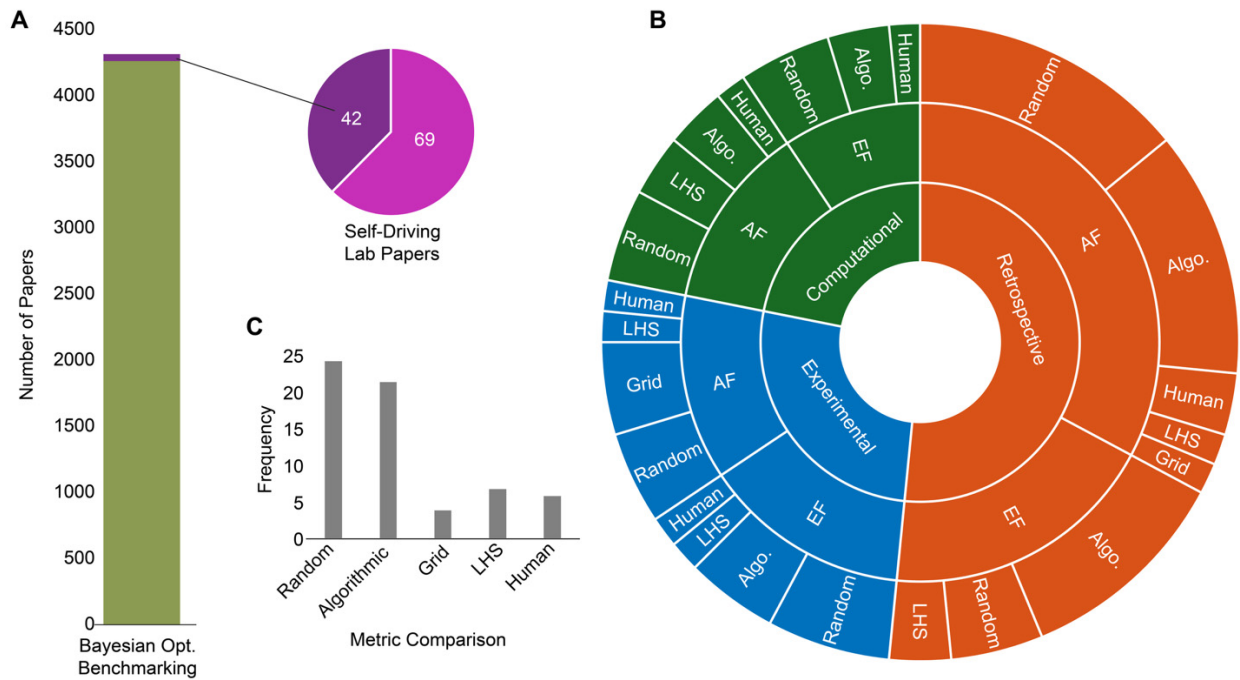

**Fig. 2** Trends in SDL benchmarking studies: (A) Summary of the Bayesian optimization benchmarking studies. The pie chart details the studies that involve SDLs. (B) Sunburst diagram depicting benchmarking results from SDL studies. The inner ring depicts the benchmarking type (experimental, retrospective, and computational), the middle ring describes the reported metric, and the outer ring depicts the reference campaign (random sampling, Latin hypercube sampling - LHS, grid-based sampling, human-directed sampling, or algorithmic to reflect a different active learning metric than Bayesian optimization). (C) Bar chart showing the number of SDL benchmarking studies that utilize each type of comparison.

Having narrowed down the field to a targeted set of papers considering experimental materials data, we set out to more fully compare this subset of the literature. The reviewed literature spans a diverse range of material domains, including electrochemistry,[19, 21-25] bulk materials discovery,[21, 26-33] spectroscopy and imaging,[21, 34] mechanics,[35-39] nanoparticle and quantum dot synthesis,[40-45] and solar cell or device optimization.[21, 30-32, 46, 47] This diversity underscores the breadth of SDL applications and highlights the variety of experimental contexts in which *AF* and *EF* are reported.

### 3.1 The Source of the Data

Benchmarking can be categorized by the source of the data, which falls into three categories.[19-24, 26-38, 40-46, 48-58] **Experimental** benchmarking are studies that complete at least two independent campaigns of experiments comparing an AL strategy to a reference strategy using unique physical experiments. This is the most informative class of benchmarking as it captures both statistical and systematic sources of experimental variability. However, this may be impractical, as it requires additional experiments that can be resource-intensive or beyond the scope of a materials study. A more attainable category of benchmarking is **retrospective**, where tables of previously completed experiments are used as ground truth for simulated campaigns. This approach has the advantages of being faster and less resource-intensive while also featuring known optima. However, decision-making policies are forced to become discrete to align with the existing data, the parameter space is vastly constricted, and noise becomes embedded into the system. Nevertheless, this approach is popular as a method to tune hyperparameters and compare algorithms. **Computational** benchmarking comprises running a campaign that queries an analytical function or computational model. This process can be fast, inexpensive, and the optima can be known for analytical functions. As such, these are extremely common in materials science and the broader optimization community for benchmarking AL algorithms. Here, we choose not to include benchmarking based on purely analytical functions and instead focus on studies that use data relevant to materials experiments, as these will provide the most direct articulation of the acceleration inherent to SDLs in materials research.

Retrospective analysis is the most common type of SDL benchmarking (Fig. 2B). For instance, Rohr *et al.* used a dataset of 2,121 catalyst compositions collected using high-throughput experimentation spanning a six-dimensional electrocatalytic metal oxide space to benchmark various sequential learning models evaluated such as Gaussian process (GP), random forest (RF), and least-squares estimation (LE).[19] The analysis, which was conducted over 1,000 learning cycles, revealed up to a 20-fold reduction in the number of experiments required to find top-performing oxygen evolution reaction catalysts, comparing GP to random sampling. The study also evaluated the effect of exploration-exploitation tuning and dataset type on model performance. Similarly, Liu *et al.* developed an SDL to optimize the open-air perovskite solar cell manufacturing process and benchmarked its BO framework using a regression model trained on experimental data.[32] They ran 300 iteration steps comparing standard BO and BO with knowledge constraint against Latin hypercube sampling (LHS), factorial sampling with progressive grid subdivision (FS-PGS), and one-variable-at-a-time sampling (OVATS). The BO methods consistently outperformed the others, showing up to a 10-fold enhancement in power conversion efficiency relative to LHS and FS-PGS.

Experimental benchmarks, while less common, are the most representative of real-world variability and experimental constraints. For example, Liu *et al.* had a limited budget of less than 100 process conditions, which limited experimental benchmarking to only standard BO vs. LHS.[32] Within 85 process conditions, BO identified four times as many high-performing perovskite films as LHS. As a separate example, Wu *et al.* benchmarked the efficiency of a BO-guided gold tetrapod nanoparticle synthesis against random search over an experimental run of 30 iterations. The BO algorithm utilized in this work, Gryffin, uses a Bayesian neural network to construct a kernel regression surrogate model. The algorithm was benchmarked based on four hierarchical objectives related to the plasmonic response of the particles. While random sampling occasionally satisfied three of the objectives, it failed to meet the final objective within the experimental budget.

Computational analyses, although sampled more selectively in this review due to our focus on benchmarking strategies that use experimental data, remain a valuable tool for comparing algorithmic strategies. Jiang *et al.* developed a chemical synthesis robot, AI-EDISON, for gold and silver nanoparticle synthesis with the goal of optimizing their optical properties.[42] As part of their workflow, they benchmarked AI-EDISON against random search in a simulated chemical space using PyDScat-GPU, a simulation tool based on discrete dipole approximation-based simulations. During a campaign with 200 steps, the algorithm outperformed random search by the 27th step, identifying samples from nine of ten spectral classes and completing all ten by the 78th step. In terms of mean fitness, which measures the similarity of a sample's spectrum to the target, AI-EDISON reached the performance achieved by 200 random steps in just 25 iterations guided by the algorithm. Annevelink *et al.* likewise developed a framework for electrochemical systems, AutoMAT, with input generation from atomic descriptors to continuum device simulations such as PyBaMM.[22] Compared to random search, AutoMAT found top-performing Li-metal electrolytes and nitrogen reduction reaction catalysts in 3 and 15 times fewer iterations respectively.

### 3.2 The Nature of the Reference Campaign

A central consideration when benchmarking learning is deciding how to select experiments for the reference campaign. We highlight the four most used reference methods. **Random sampling** involves choosing each experiment uniformly at random in the parameter space. Random sampling is simple to implement and will converge in a predictable manner, as described by Eq. (1). Furthermore, the total number of experiments does not have to be chosen prior to the campaign, which facilitates analysis and data reuse. **Grid-based sampling** involves dividing the parameter space into uniformly spaced intervals. It is easy to implement and will provide a balanced view across parameter space, but at the cost of needing to specify the total number of experiments *a priori*. **Latin hypercube sampling (LHS)** combines the even distribution of grid sampling with the perturbations of random sampling to provide a balanced picture of parameter space while using any number of points. This is generally the preferred method for obtaining data when performing initial training campaigns. Like grid sampling, an LHS campaign cannot be stopped early without having a biased data distribution and relying on evenly distributed samples may over-sample flat regions while potentially missing areas with sharp transitions. **Human-directed sampling** is the non-SDL state of the art and provides a useful comparison when evaluating whether the algorithm is providing value. However, human-directed sampling is time-consuming and introduces variability and bias from individual decision-making. All four of these methods have been explored for benchmarking (Fig. 2C).

Across the reviewed SDL papers, which include 42 unique studies and 63 reported benchmarks, the most fundamental and widely adopted baseline is random sampling. MacLeod *et al.* evaluated their SDL, Ada, for multi-objective optimization of palladium film synthesis, balancing conductivity and annealing temperature.[46] In a simulated campaign using a model built from experimental data, Ada's q-expected hypervolume improvement (q-EHVI) strategy achieved twice the hypervolume of random sampling within 25 steps and reached a hypervolume achieved by 10,000 random samples in just 100 steps. Similarly, Bai *et al.* developed a platform to explore the copper antimony sulfide (Cu-Sb-S) compositional space for photo-electrocatalytic hydrogen evolution. In this experimental benchmarking study, the Bayesian optimizer revealed a Cu-Sb-S composition that exhibited 2.3 times greater catalytic activity than results from random sampling.

Many SDL studies compare performance between algorithms, which frequently includes variants of BO (e.g., differing surrogate models, acquisition functions, or kernels),[49] as well as hybridized approaches involving evolutionary algorithms,[29, 46] or reinforcement learning.[43] For instance, Ziomek *et al.* proposed a length scale balancing GP-UCB (LB-GP-UCB), a BO variant with an upper confidence bound (UCB) acquisition function that aggregates multiple GPs with different length scales to address the challenge of unknown kernel hyperparameters.[39] It retrospectively benchmarked the performance of LB-GP-UCB against adaptive GP-UCB (A-GP-UCB),[59] maximum likelihood estimation (MLE),[60] and Markov chain Monte Carlo (MCMC)[61] using the crossed barrel[35] and silver nanoparticle[62] datasets. For both datasets, LB-GP-UCB consistently found the optimal solution with fewer experiments, specifically requiring 40% fewer trials than MLE and MCMC.

A relatively small number of studies reported performance relative to LHS and grid-based sampling. Gongora *et al.* developed the Bayesian experimental autonomous researcher (BEAR) to optimize the toughness of crossed barrel structures.[35, 57] They benchmarked its performance against grid sampling, where the 4D design space was discretized into 600 points, each tested in triplicate. The BEAR running on a BO framework with an expected improvement (EI) acquisition function discovered higher-performing structures with 18 times fewer experiments. Also, Bateni *et al.* developed an SDL, Smart Dope, for space exploration and optimization of lead halide perovskite (LHP) quantum dots (QDs).[40] Using LHS, 150 initial experiments were conducted across the nine-dimensional space to generate training data for closed-loop optimization. Smart Dope, also running on BO with an expected improvement acquisition function, achieved a photoluminescence quantum yield (PLQY) of 158% after just four closed-loop iterations, exceeding the 151% maximum obtained by LHS. This suggests that LHS and grid-based sampling's fixed intervals may over-represent flat regions while missing sharp transitions.

Human-directed sampling, where expert researchers select experimental conditions based on intuition and domain knowledge, also appears in the reviewed SDL literature, and it provides a useful comparison between SDLs and conventional experimentation. Nakayama *et al.* benchmarked BO against human-directed sampling using a one-dimensional model of synthesis temperature optimization.[48] Human experts required 13-14 trials to find the global optimum, while BO required only ten steps with the appropriate acquisition function and hyperparameters. The search efficiency of BO demonstrated in this simple 1D case will grow in higher-dimensional spaces where human intuition is more limited. Sheilds *et al.* benchmarked the performance of BO against 50 expert chemists using high-throughput experimental data covering a ten-dimensional parameter space for optimizing the yield of direct arylation of imidazoles.[56] To reduce bias, the performance was averaged across the 50 human participants and 50 runs of the Bayesian optimizer, each conducted over 100 steps. While humans achieved 15% higher yield in the first five experiments, by the 15$^{th}$ experiment, the average performance of the optimizer surpassed that of the humans. BO consistently achieved >99% yield within the experimental budget, and within the first 50 experiments, it discovered the global optimum that none of the experts found.

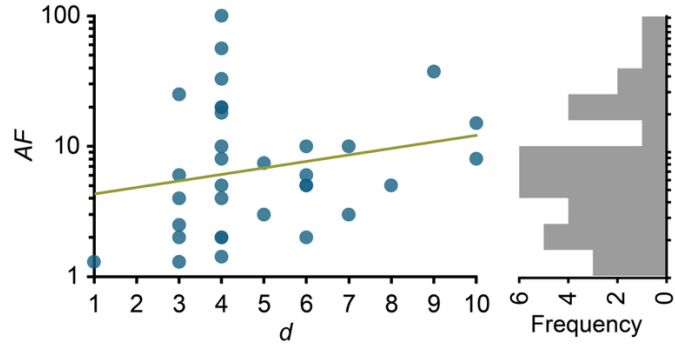

**Fig. 3** Acceleration factor ($AF$) vs. parameter space dimensionality $d$ across benchmarking SDL studies, with corresponding $AF$ frequency.

**Table 1.** Summary of reported $AF$ from SDL benchmarking studies

| Case | Source | $AF$ | Type | Dimension | Comparison | Objective |
|---|---|---|---|---|---|---|
| 1 | Bateni et al.[40] | 37.5 | Experimental | 9 | GP-EI vs. LHS | Photoluminescence quantum yield |
| 2 | Cakan et al.[30] | 2.5 | Experimental | 3 | GP-EI vs. grid | Film photothermal stability |
| 3 | Fatehi et al.[25] | 20 | Experimental | 4 | GP-EI & GP-UCB vs. random search | Catalyst activity |
| 4 | Gongora et al.[35] | 18 | Experimental | 4 | GP-EI vs. grid (best grid performance as reference) | Structure toughness |
| 5 | Gongora et al.[35] | 56.25 | Experimental | 4 | GP-EI vs. grid (best BO performance within a time budget as reference) | Structure toughness |
| 6 | Gongora et al.[36] | 10 | Experimental | 4 | GP-EI (FEA informed) vs. GP-EI (uninformative prior) | Structure toughness |
| 7 | Wu et al.[44] | 10 | Experimental | 7 | Gryffin algorithm (BO based on kernel density estimation) vs. random search | Nanoparticle plasmonic response |
| 8 | Borg et al.[26] | 2 | Retrospective | 3 | RF-EI & RF-EV (expected value) vs. random search (identifying single target material) | Band gap of inorganics |
| 9 | Borg et al.[26] | 4 | Retrospective | 3 | RF-EI & RF-EV vs. random search (identifying five target materials) | Band gap of inorganics |
| 10 | Dave et al.[23] | 1.3 | Retrospective | 3 | Random search vs. human | Electrolyte ionic conductivity |
| 11 | Dave et al.[23] | 6 | Retrospective | 3 | GP-MLE vs. random search | Electrolyte ionic conductivity |
| 12 | Guay-Hottin et al.[49] | 1.42 | Retrospective | 4 | α-πBO (GP-EI with dynamic hyperparameter tuning) vs. standard GP-EI | Structure toughness |
| 13 | Langner et al.[31] | 33 | Retrospective | 4 | Bayesian neural network (BNN) vs. grid | Film photostability |
| 14 | Liang et al.[20] | 2 | Retrospective | 4 | GP-ARD (automatic relevance detection)-LCB vs. random search | Structure toughness |
| 15 | Liang et al.[20] | 8 | Retrospective | 4 | RF-LCB (lower confidence bound) vs. random search | Structure toughness |
| 16 | Liang et al.[20] | 4 | Retrospective | 4 | GP-LCB (lower confidence bound) vs. random search | Structure toughness |
| 17 | Liu et al.[32] | 61 | Retrospective | 6 | Standard BO & knowledge-constrained BO vs. LHS | Film power conversion efficiency |
| 18 | Lookman et al.[28] | 3 | Retrospective | 7 | GP-EI vs. random search | Material electrostrain |

| 19 | Low et al.[29] | 5 | Retrospective | 8 | qNEHVI (q-noisy expected hypervolume improvement) vs. U-NSGA-III (unified non-dominated sorting genetic algorithm III) | Concrete slump & compressive strength |
| 20 | Low et al.[29] | 20 | Retrospective | 4 | qNEHVI vs. U-NSGA-III | Film conductivity & annealing temperature |
| 21 | MacLeod et al.[46] | 100 | Retrospective | 4 | qEHVI (q-expected hypervolume improvement) vs. random search | Film conductivity & annealing temperature |
| 22 | Rohr et al.[19] | 10 | Retrospective | 6 | RF-UCB & GP-UCB vs. random search | Catalyst activity |
| 23 | Rohr et al.[19] | 5 | Retrospective | 6 | LE (linear ensemble) vs. random search | Catalyst activity |
| 24 | Ros et al.[33] | 5 | Retrospective | 6 | GP-EI-Thompson sampling & vs. random search | Drug solubility |
| 25 | Thelen et al.[24] | 5 | Retrospective | 4 | GP-EI & GP-PI (probability of improvement) vs. random search | Battery cycle life |
| 26 | Thelen et al.[24] | 2 | Retrospective | 4 | GP-UCB vs. random search | Battery cycle life |
| 27 | Ament et al.[21] | 25 | Computational | 3 | GP-IGU (integrated gradient uncertainty) vs. random search | Phase boundary mapping |
| 28 | Annevelink et al.[22] | 3 | Computational | 5 | AutoMat-FUELS (forests with uncertainty estimates for learning sequentially) vs. random search | Catalyst activity |
| 29 | Annevelink et al.[22] | 15 | Computational | 10 | AutoMat-FUELS vs. random search | Battery cycle life |
| 30 | Jiang et al.[42] | 7.41 | Computational | 5 | Quality diversity (QD) algorithm vs. random search | Nanoparticle extinction spectra |
| 31 | Lei et al.[57] | 8 | Computational | 10 | BART (Bayesian additive regression trees) & BMARS (Bayesian multivariate adaptive regression splines) vs. standard BO | Crystal stacking fault energy |
| 32 | Lookman et al.[28] | 2 | Computational | 6 | GP-EI vs. RF + EI | LED quantum efficiency |
| 33 | Nakayama et al.[48] | 1.3 | Computational | 1 | GP-EI vs. human | Synthesis temperature |

## 3.3 Meta Analysis of Reported Benchmarking

To visualize the reported SDL benchmarking, we extracted $AF$ from studies spanning a range of $d$ (Fig. 3). Overall, the reported $AF$ spanned a wide range, from 1.3 to 100, highlighting the variability in how effectively active learning accelerates research across different experimental domains. The median reported $AF$ was 6. Interestingly, $AF$ appeared to increase with increasing $d$, suggesting that the "curse of dimensionality" was managed more effectively by active learning than by random sampling. From a learning efficiency perspective, this suggests a "blessing of dimensionality" in which higher-dimensional spaces provide more incentive to use advanced learning algorithms. A summary of the $AF$ values is provided in Table 1. To provide some notable examples, at the low end, an $AF$ of 1.3 was observed in a 1D temperature-dependent synthesis optimization task, where the number of iterations required for BO to locate the global maximum was compared to that required by a human researcher.[48] At the high end, a multi-objective Bayesian optimization campaign for metallic thin-film synthesis in a 4D parameter space achieved an $AF$ of 100 when benchmarked against random sampling.[46]

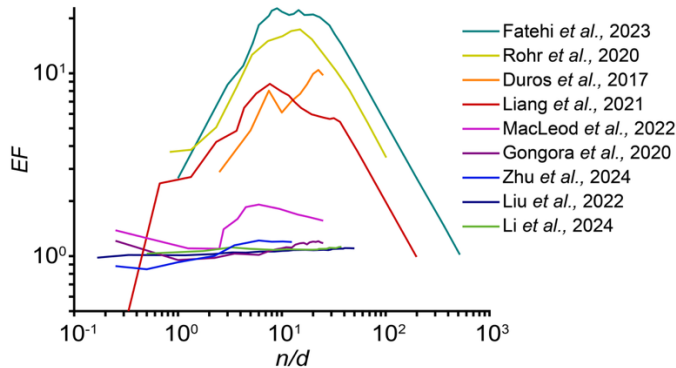

**Fig. 4** $EF$ vs. experiment number $n$ normalized by $d$, extracted from performance-over-iteration data (relative to random sampling) in experimental and retrospective benchmarking SDL studies

While $AF$ is simple to report, it is subtle to interpret as it depends on the chosen performance threshold. Typically, this threshold corresponds either to a value defined by the researcher or the highest performance achieved during the campaign.[30, 46] In contrast, $EF$ is easy to calculate at each experiment, and it does not rely on a performance value, making it useful for tracking learning progress.

In order to visualize $EF$ progression over the course of SDL campaigns, we extracted $EF$ from reported performance trajectories (Fig. 4). We limited this analysis to studies that benchmarked against random sampling since this can serve as a common baseline. To enable comparison across studies with different $d$, we divided experiment number $n$ by $d$. We focused specifically on experimental and retrospective benchmarking studies, as these are grounded in real experimental data. Examining the computed $EF$ values, a consistent pattern emerges in which $EF$ initially grows with $n/d$, reaches a peak, and then gradually declines. This indicates that the benefit from active learning is most important early in a campaign, where the algorithm can make rapid progress towards the chosen goal. At higher numbers of experiments, the diminishing marginal gains of active learning combined with the continual progress of random sampling mean that the benefit of active learning becomes less important. In other words, if enough of the parameter space will be sampled, the order in which it is sampled is not important. Interestingly, this peak in $EF$ occurs at ~10 to 20 experiments per dimension, which provides a useful reference point for the SDL community when planning campaigns.

While the number of experiments at which $EF$ peaked was relatively consistent, the peak value of $EF$ varied substantially between studies. The analysis in Section 2 reveals that the maximum attainable value for $EF$ is $C$, which depends on the property space. For example, the largest $EF$ observed in our analysis was 23, reported by Fatehi et al.,[25] who applied a Bayesian optimization framework with a UCB acquisition function to quantify the proportion of top-performing oxygen evolution reaction (OER) catalysts identified relative to random sampling, using the dataset by Rohr et al.[19] In contrast, Zhu et al.[38] using experimental design via Bayesian optimization package (EDBO)[56] and Li et al.[37] using graph-based Bayesian optimization with pseudo labeling (GBOPL), both benchmarked their algorithms on the crossed barrel dataset, to find modest maximum $EF$ of 1.2 and 1.1, reflecting the narrower performance gap in this property space. This is similar to the $EF$ of 1.2 observed in the experimental benchmarking study by Gongora *et al.*,[35] the source of the dataset.

## 4. Exploration of Benchmarking Metrics

While it is clear from the reported values of $EF$ that this metric varies dramatically, it is not clear how this should be interpreted or whether this variation is due to differences in algorithms or the underlying parameter spaces. To explore this, we perform a series of simulated Bayesian optimization campaigns designed to illuminate how $EF(n)$ depends on the underlying parameter space. In particular, we develop a simple two-dimensional parameter space that features a single Gaussian peak in the center of the space (Fig. 5A). The results of simulated BO campaigns in this space are reported as a horse race plot in which shaded regions depict the quartile ranges from 100 independent campaigns (Fig. 5B). These are compared to campaigns based on sampling uniformly at random which center on the theoretical performance predicted by Equation (1). These campaigns were executed using the BoTorch package, and the code is shared at https://github.com/kabrownlab/benchmarking.

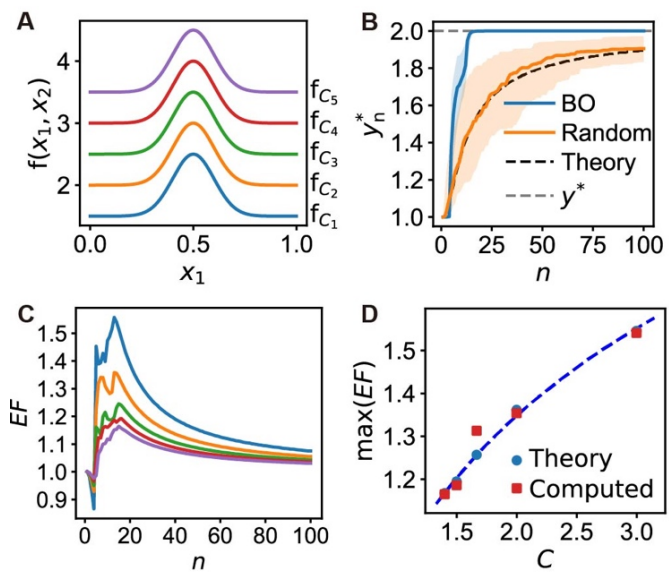

**Fig. 5** Simulated Bayesian optimization (BO) campaigns to explore how the property space dictates convergence. (A) Five two-dimensional functions $f$ under consideration that differ only in their contrast $C = \max(f)/\text{median}(f)$. While all are two-dimensional, they depend on $x_1$ and $x_2$ in the same way and $x_2 = 0.5$ is shown. (B) Simulated horse race plot showing the convergence of BO and random sampling. Theory corresponds to Eq. (1). The shaded regions show interquartile ranges. (C) $EF$ vs. $n$ for the five functions shown in (A). (D) $\max(EF)$ relating BO and random sampling vs. $C$. Dashed line shows a fit to $\max(EF) = a \log C + b$.

In a first round of simulations to explore the magnitude of $\max(EF)$, we performed optimization campaigns using five functions that differed only in their contrast $C$ (Fig. 5A). As expected, all campaigns achieved a $\max(EF)$ at similar $n$ but exhibited very different magnitudes depending on the function (Fig. 5C). Indeed, the theoretical and computed $\max(EF)$ followed identical trends and monotonically increased with $C$ (Fig. 5D). This analysis confirms that while the complexity of the function dictates how many samples are needed to find an optima, its $C$ bounds $EF$, explaining why the literature features such a wide range in reported $\max(EF)$.

While the functions explored in Fig. 5 exhibited the same complexity, we sought to explore whether one can use simple statistics of a function to gain insight into how many experiments are needed to achieve optimum performance. In particular, we explore Lipschitz complexity $L$, which is defined as,[63]

$$L = \max|\nabla f|, \tag{5}$$

where $|\nabla f|$ represents the magnitude of the gradient of the function $f$ in which each independent variable has been normalized to fall between 0 and 1. We construct a family of functions with the same $C$ but different $L$ by changing the width of two-dimensional Gaussians (Fig. 6A). Unlike the case when only $C$ is changed, each campaign requires different numbers of experiments to converge with sharper functions requiring more experiments (Fig. 6B). Interestingly, we find a linear relationship between $L$ and $n_{AL}$, highlighting the challenge inherent to parameter spaces that appear to be needles in a haystack. Interestingly, the empirically observed best experiment number $n_{AL}^*$ from the literature appears to be ~15/$d$, which amounts to 30 experiments in the present example. This suggests that the functions explored here share statistical features with the materials spaces previously studied. Importantly, $\max(EF)$ increases with $L$, highlighting that it is more impactful to use active learning in parameter spaces that are more difficult to learn.

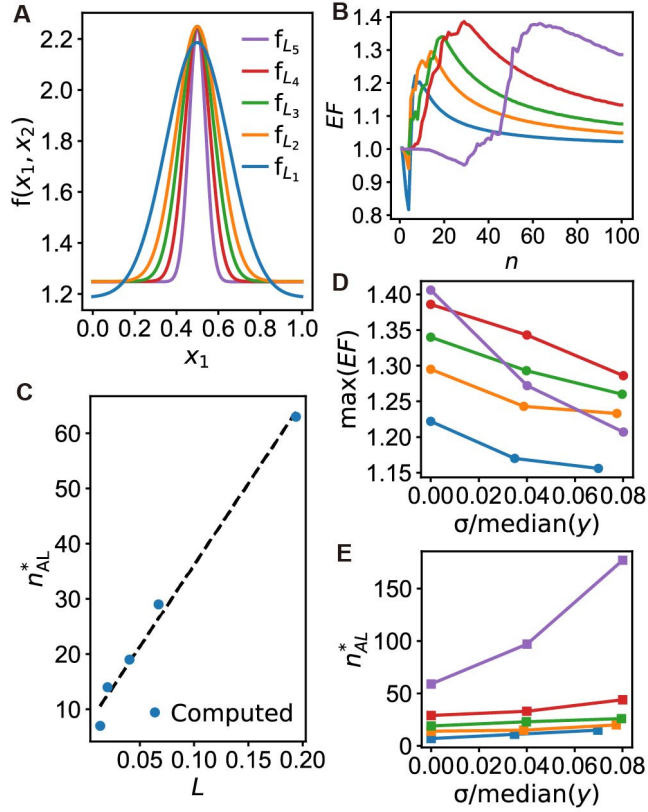

**Fig. 6** Simulated BO campaigns to explore how property space complexity impacts learning. (A) Five two-dimensional parameter spaces $f$ under consideration that differ only in their Lipschitz complexity $L$, as defined in Eq. (5). While all are two-dimensional, they depend on $x_1$ and $x_2$ in the same manner and $x_2 = 0.5$ is shown. (B) $EF$ vs. $n$ for the five functions shown in (A). (C) Optimum experiment number $n_{AL}^*$ corresponding to $\max(EF)$ vs. $L$. The dashed line shows a linear fit. (D) $\max(EF)$ vs. noise standard deviation $\sigma$ normalized by $\mathrm{median}(y)$. (E) $n_{AL}^*$ and vs. $\sigma$ normalized by $\mathrm{median}(y)$.

The analytical spaces considered here are deterministic, while experimental parameter spaces will necessarily feature noise. In an effort to understand how the presence of noise will impact convergence, simulated BO campaigns were repeated for the functions shown in Fig. 6A with homoscedastic Gaussian noise with standard deviation $\sigma$ added. While $\max(EF)$ had a weak and smooth dependence on $\sigma$ (Fig. 6D), $n_{AL}^*$ depended sharply on $\sigma$, with the most complex functions exhibiting drastic increases in $n_{AL}^*$ (Fig. 6E). This result indicates that reducing noise becomes more important the more complex the parameter space.

# 5. Conclusions and Future Recommendations

Benchmarking SDLs is important because it provides part of the justification for developing and running these systems. As a result, there have been significant efforts in the community to quantify performance. The two most reported metrics are the enhancement factor $EF$ and the acceleration factor $AF$, which address the questions of how much better and how much faster, respectively. A systematic evaluation of the reported metrics reveals key insights:

i. SDLs achieve top-performing results on average six times faster than random sampling, and this acceleration improves with the dimensionality of the parameter space.
ii. The enhancement inherent to SDLs is reported to peak at 10-20 experiments per dimension of parameter space, with enhancement factors that vary tremendously depending on the space.

It is important to highlight that both of these outcomes depend intimately on the nature of the property spaces, but the fact that these all represent actual experimental materials datasets suggests that they are useful guidelines for the field. Further, simulated campaigns in analytical spaces reveal key features of how to interpret metrics, namely that $EF$ can simply be related to the statistics of the parameter space such as its contrast, the complexity of the space determines the speed with which convergence can be expected, and that noise affects $AF$ more than $EF$. While the specific values in this study will hopefully be improved upon in the coming years as more advanced algorithms are employed, they nevertheless provide a valuable snapshot of the field and a useful tool to align progress. Addressing the materials challenges facing our society demands rapid progress and a thorough analysis of methods to accelerate this progress is necessary to move the field forward.

**Acknowledgements**

The authors thank Robert Brown and Peter Frazier for helpful conversations. The authors acknowledge the Hariri Institute at Boston University (2024-07-001), the National Science Foundation (DMR-2323728), and the Army Research Office (W911NF2420095) for supporting this research.

**Conflict of Interest Statement**

The authors declare no conflicts of interest.

**Data Availability Statement**

The simulations used to create the plots in Figures 5 and 6 and the data used to create the plots in Figures 3 and 4 are available at: https://github.com/kabrownlab/benchmarking

# References

1. J. J. de Pablo, N. E. Jackson, M. A. Webb, L.-Q. Chen, J. E. Moore, D. Morgan, R. Jacobs, T. Pollock, D. G. Schlom and E. S. Toberer, *Npj Comput Mater*, 2019, **5**, 1-23.
2. R. B. Canty, J. A. Bennett, K. A. Brown, T. Buonassisi, S. V. Kalinin, J. R. Kitchin, B. Maruyama, R. G. Moore, J. Schrier, M. Seifrid, S. Sun, T. Vegge and M. Abolhasani, *Nat Commun*, 2025, **16**, 3856.
3. R. Vescovi, T. Ginsburg, K. Hippe, D. Ozgulbas, C. Stone, A. Stroka, R. Butler, B. Blaiszik, T. Brettin, K. Chard, M. Hereld, A. Ramanathan, R. Stevens, A. Vriza, J. Xu, Q. Zhang and I. Foster, *Digit Discov*, 2023, **2**, 1980-1998.
4. E. Stach, B. DeCost, A. G. Kusne, J. Hattrick-Simpers, K. A. Brown, K. G. Reyes, J. Schrier, S. Billinge, T. Buonassisi, I. Foster, C. P. Gomes, J. M. Gregoire, A. Mehta, J. Montoya, E. Olivetti, C. Park, E. Rotenberg, S. K. Saikin, S. Smullin, V. Stanev and B. Maruyama, *Matter*, 2021, **4**, 2702–2726.
5. G. Tom, S. P. Schmid, S. G. Baird, Y. Cao, K. Darvish, H. Hao, S. Lo, S. Pablo-García, E. M. Rajaonson, M. Skreta, N. Yoshikawa, S. Corapi, G. D. Akkoc, F. Strieth-Kalthoff, M. Seifrid and A. Aspuru-Guzik, *Chemical Reviews*, 2024, **124**, 9633-9732.
6. E. J. Kluender, J. L. Hedrick, K. A. Brown, R. Rao, B. Meckes, J. Du, L. Moreau, B. Maruyama and C. A. Mirkin, *Proceedings of the National Academy of Sciences of the United States of America*, 2019, **116** 40–45.
7. J. Zhou, M. Luo, L. Chen, Q. Zhu, S. Jiang, F. Zhang, W. Shang and J. Jiang, *Digit Discov*, 2025, **4**, 636-652.
8. M. Zaki, C. Prinz and B. Ruehle, *Acs Nano*, 2025, DOI: 10.1021/acsnano.4c17504.
9. C. Wang, Y.-J. Kim, A. Vriza, R. Batra, A. Baskaran, N. Shan, N. Li, P. Darancet, L. Ward, Y. Liu, M. K. Y. Chan, S. K. R. S. Sankaranarayanan, H. C. Fry, C. S. Miller, H. Chan and J. Xu, *Nat Commun*, 2025, **16**, 1498.
10. T. Song, M. Luo, X. Zhang, L. Chen, Y. Huang, J. Cao, Q. Zhu, D. Liu, B. Zhang, G. Zou, G. Zhang, F. Zhang, W. Shang, Y. Fu, J. Jiang and Y. Luo, *Journal of the American Chemical Society*, 2025, **147**, 12534-12545.
11. A. Sanin, J. K. Flowers, T. H. Piotrowiak, F. Felsen, L. Merker, A. Ludwig, D. Bresser and H. S. Stein, *Adv Energy Mater*, 2025, **15**, 2404961.
12. S. Putz, J. Döttling, T. Ballweg, A. Tschöpe, V. Biniyaminov and M. Franzreb, *Advanced Intelligent Systems*, 2025, **7**, 2400564.
13. K. Nishio, A. Aiba, K. Takihara, Y. Suzuki, R. Nakayama, S. Kobayashi, A. Abe, H. Baba, S. Katagiri, K. Omoto, K. Ito, R. Shimizu and T. Hitosugi, *Digit Discov*, 2025, **4**, 1734-1742.
14. F. Strieth-Kalthoff, H. Hao, V. Rathore, J. Derasp, T. Gaudin, N. H. Angello, M. Seifrid, E. Trushina, M. Guy, J. Liu, X. Tang, M. Mamada, W. Wang, T. Tsagaantsooj, C. Lavigne, R. Pollice, T. C. Wu, K. Hotta, L. Bodo, S. Li, M. Haddadnia, A. Wołos, R. Roszak, C. T. Ser, C. Bozal-Ginesta, R. J. Hickman, J. Vestfrid, A. Aguilar-Granda, E. L. Klimareva, R. C. Sigerson, W. Hou, D. Gahler, S. Lach, A. Warzybok, O. Borodin, S. Rohrbach, B. Sanchez-Lengeling, C. Adachi, B. A. Grzybowski, L. Cronin, J. E. Hein, M. D. Burke and A. Aspuru-Guzik, *Science*, 2024, **384**, eadk9227.
15. K. L. Snapp, B. Verdier, A. E. Gongora, S. Silverman, A. D. Adesiji, E. F. Morgan, T. J. Lawton, E. Whiting and K. A. Brown, *Nat Commun*, 2024, **15**, 4290.
16. S. Matsuda, G. Lambard and K. Sodeyama, *Cell Reports Physical Science*, 2022, **3**, 100832.
17. F. Delgado-Licona and M. Abolhasani, *Advanced Intelligent Systems*, 2023, **5**, 2200331.
18. L. Hung, J. A. Yager, D. Monteverde, D. Baiocchi, H.-K. Kwon, S. Sun and S. Suram, *Digit Discov*, 2024, **3**, 1273-1279.
19. B. Rohr, H. S. Stein, D. Guevarra, Y. Wang, J. A. Haber, M. Aykol, S. K. Suram and J. M. Gregoire, *Chem. Sci.*, 2020, **11**, 2696-2706.

20. Q. H. Liang, A. E. Gongora, Z. K. Ren, A. Tiihonen, Z. Liu, S. J. Sun, J. R. Deneault, D. Bash, F. Mekki-Berrada, S. A. Khan, K. Hippalgaonkar, B. Maruyama, K. A. Brown, J. I. I. I. Fisher and T. Buonassisi, *Npj Comput Mater*, 2021, **7**.
21. S. Ament, M. Amsler, D. R. Sutherland, M. C. Chang, D. Guevarra, A. B. Connolly, J. M. Gregoire, M. O. Thompson, C. P. Gomes and R. B. van Dover, *Sci Adv*, 2021, **7**.
22. E. Annevelink, R. Kurchin, E. Muckley, L. Kavalsky, V. I. Hegde, V. Sulzer, S. Zhu, J. K. Pu, D. Farina, M. Johnson, D. Gandhi, A. Dave, H. Y. Lin, A. Edelman, B. Ramsundar, J. Saal, C. Rackauckas, V. R. Shah, B. Meredig and V. Viswanathan, *Mrs Bull*, 2022, **47**, 1036-1044.
23. A. Dave, J. Mitchell, S. Burke, H. Y. Lin, J. Whitacre and V. Viswanathan, *Nat Commun*, 2022, **13**.
24. A. Thelen, M. Zohair, J. Ramamurthy, A. Harkaway, W. M. Jiao, M. Ojha, M. Ul Ishtiaque, T. A. Kingston, C. L. Pint and C. Hu, *J Power Sources*, 2023, **581**.
25. E. Fatehi, M. Thadani, G. Birsan and R. W. Black, *arXiv preprint arXiv:2305.12541*, 2023.
26. C. K. H. Borg, E. S. Muckley, C. Nyby, J. E. Saal, L. Ward, A. Mehta and B. Meredig, *Digit Discov*, 2023, **2**, 327-338.
27. P. Honarmandi, V. Attari and R. Arroyave, *Comp Mater Sci*, 2022, **210**.
28. T. Lookman, P. V. Balachandran, D. Z. Xue and R. H. Yuan, *Npj Comput Mater*, 2019, **5**.
29. A. K. Y. Low, E. Vissol-Gaudin, Y. F. Lim and K. Hippalgaonkar, *J Mater Inform*, 2023, **3**.
30. D. N. Cakan, E. Oberholtz, K. Kaushal, S. P. Dunfield and D. P. Fenning, *Mater Adv*, 2025, **6**, 598-606.
31. S. Langner, F. Häse, J. D. Perea, T. Stubhan, J. Hauch, L. M. Roch, T. Heumueller, A. Aspuru-Guzik and C. J. Brabec, *Adv Mater*, 2020, **32**.
32. Z. Liu, N. Rolston, A. C. Flick, T. W. Colburn, Z. K. Ren, R. H. Dauskardt and T. Buonassisi, *Joule*, 2022, **6**, 834-849.
33. H. Ros, M. Cook and D. Shorthouse, 2024.
34. R. K. Vasudevan, K. P. Kelley, J. Hinkle, H. Funakubo, S. Jesse, S. V. Kalinin and M. Ziatdinov, *Acs Nano*, 2021, **15**, 11253-11262.
35. A. E. Gongora, B. W. Xu, W. Perry, C. Okoye, P. Riley, K. G. Reyes, E. F. Morgan and K. A. Brown, *Sci Adv*, 2020, **6**.
36. A. E. Gongora, K. L. Snapp, E. Whiting, P. Riley, K. G. Reyes, E. F. Morgan and K. A. Brown, *Iscience*, 2021, **24**.
37. G. Y. Li and X. N. Jin, *Ieee Int Con Auto Sc*, 2024, DOI: 10.1109/Case59546.2024.10711759, 2955-2960.
38. M. J. Zhu, A. Mroz, L. F. Gui, K. E. Jelfs, A. Bemporad, E. A. D. Chanona and Y. S. Lee, *Digit Discov*, 2024, **3**, 2589-2606.
39. J. Ziomek, M. Adachi and M. A. Osborne, *arXiv preprint arXiv:2410.10384*, 2024.
40. F. Bateni, S. Sadeghi, N. Orouji, J. A. Bennett, V. S. Punati, C. Stark, J. Y. Wang, M. C. Rosko, O. Chen, F. N. Castellano, K. G. Reyes and M. Abolhasani, *Adv Energy Mater*, 2024, **14**.
41. R. W. Epps, M. S. Bowen, A. A. Volk, K. Abdel-Latif, S. Y. Han, K. G. Reyes, A. Amassian and M. Abolhasani, *Adv Mater*, 2020, **32**.
42. Y. B. Jiang, D. Salley, A. Sharma, G. Keenan, M. Mullin and L. Cronin, *Sci Adv*, 2022, **8**.
43. A. A. Volk, R. W. Epps, D. T. Yonemoto, B. S. Masters, F. N. Castellano, K. G. Reyes and M. Abolhasani, *Nat Commun*, 2023, **14**.
44. T. Y. Wu, S. Kheiri, R. J. Hickman, H. C. Tao, T. C. Wu, Z. B. Yang, X. Ge, W. Zhang, M. Abolhasani, K. Liu, A. Aspuru-Guzik and E. Kumacheva, *Nat Commun*, 2025, **16**.
45. S. Sadeghi, F. Bateni, T. Kim, D. Y. Son, J. A. Bennett, N. Orouji, V. S. Punati, C. Stark, T. D. Cerra, R. Awad, F. Delgado-Licona, J. E. Xu, N. Mukhin, H. Dickerson, K. G. Reyes and M. Abolhasani, *Nanoscale*, 2024, **16**, 580-591.
46. B. P. MacLeod, F. G. L. Parlane, C. C. Rupnow, K. E. Dettelbach, M. S. Elliott, T. D. Morrissey, T. H. Haley, O. Proskurin, M. B. Rooney, N. Taherimakhsousi, D. J. Dvorak, H. N. Chiu, C. E. B. Waizenegger, K. Ocean, M. Mokhtari and C. P. Berlinguette, *Nat Commun*, 2022, **13**.

47. C. Tamura, H. Job, H. Chang, W. Wang, Y. Liang and S. Sun, 2025.
48. R. Nakayama, R. Shimizu, T. Haga, T. Kimura, Y. Ando, S. Kobayashi, N. Yasuo, M. Sekijima and T. Hitosugi, *Sci Tech Adv Mat-Met*, 2022, **2**, 119-128.
49. R. Guay-Hottin, L. Kardassevitch, H. Pham, G. Lajoie and M. Bonizzato, *Knowl-Based Syst*, 2025, **311**.
50. V. Duros, J. Grizou, W. M. Xuan, Z. Hosni, D. L. Long, H. N. Miras and L. Cronin, *Angew Chem Int Edit*, 2017, **56**, 10815-10820.
51. J. Grizou, L. J. Points, A. Sharma and L. Cronin, *Sci Adv*, 2020, **6**.
52. Y. Bai, Z. H. J. Khoo, R. Made, H. Q. Xie, C. Y. J. Lim, A. D. Handoko, V. Chellappan, J. J. Cheng, F. X. Wei, Y. F. Lim and K. Hippalgaonkar, *Adv Mater*, 2024, **36**.
53. L. Kavalsky, V. I. Hegde, E. Muckley, M. S. Johnson, B. Meredig and V. Viswanathan, *Digit Discov*, 2023, **2**, 1112-1125.
54. Q. Liang, A. E. Gongora, Z. Ren, A. Tiihonen, Z. Liu, S. Sun, J. R. Deneault, D. Bash, F. Mekki-Berrada, S. A. Khan, K. Hippalgaonkar, B. Maruyama, K. A. Brown, J. Fisher Iii and T. Buonassisi, *Npj Comput Mater*, 2021, **7**, 188.
55. F. Conrad, M. Mälzer, M. Schwarzenberger, H. Wiemer and S. Ihlenfeldt, *Sci Rep-Uk*, 2022, **12**.
56. B. J. Shields, J. Stevens, J. Li, M. Parasram, F. Damani, J. I. M. Alvarado, J. M. Janey, R. P. Adams and A. G. Doyle, *Nature*, 2021, **590**, 89-96.
57. B. W. Lei, T. Q. Kirk, A. Bhattacharya, D. Pati, X. N. Qian, R. Arroyave and B. K. Mallick, *Npj Comput Mater*, 2021, **7**.
58. D. A. Cohn, Z. Ghahramani and M. I. Jordan, *J Artif Intell Res*, 1996, **4**, 129-145.
59. F. Berkenkamp, A. P. Schoellig and A. Krause, *Journal of Machine Learning Research*, 2019, **20**, 1-24.
60. D. C. Liu and J. Nocedal, *Mathematical programming*, 1989, **45**, 503-528.
61. M. D. Hoffman and A. Gelman, *J. Mach. Learn. Res.*, 2014, **15**, 1593-1623.
62. F. Mekki-Berrada, Z. Ren, T. Huang, W. K. Wong, F. Zheng, J. Xie, I. P. S. Tian, S. Jayavelu, Z. Mahfoud and D. Bash, *Npj Comput Mater*, 2021, **7**, 55.
63. G. R. Wood and B. Zhang, *Journal of Global Optimization*, 1996, **8**, 91-103.